# ANALYSIS OF THERMALLY INDUCED FREQUENCY SHIFT FOR THE SPALLATION NEUTRON SOURCE RFQ[*]

S. Virostek, J. Staples,
LBNL, Berkeley, CA 94720, USA

## Abstract

The Spallation Neutron Source (SNS) Front-End Systems Group at Lawrence Berkeley National Lab (LBNL) is developing a Radio Frequency Quadrupole (RFQ) to accelerate an H- beam from 65 keV to 2.5 MeV at the operating frequency of 402.5 MHz. The 4 section, 3.7 meter long RFQ is a 4 vane structure operating at 6% duty factor. The cavity walls are made from OFE Copper with a GlidCop® outer layer to add mechanical strength. A set of 12 cooling channels in the RFQ cavity walls are fed and controlled separately from 4 channels embedded in the vanes. An ANSYS® finite-element model has been developed to calculate the deformed shape of the cavity for given RF heat loads and cooling water temperatures. By combining the FEA results with a SUPERFISH RF cavity simulation, the relative shift in frequency for a given change in coolant temperature or heat load can be predicted. The calculated cavity frequency sensitivity is -33 kHz per 1°C change in vane water temperature with constant-temperature wall water. The system start-up transient was also studied using the previously mentioned FEA model. By controlling the RF power ramp rate and the independent wall and vane cooling circuit temperatures, the system turn-on time can be minimized while limiting the frequency shift.

## 1 INTRODUCTION

The SNS is an accelerator-based facility to be built for the US Department of Energy at Oak Ridge National Laboratory (ORNL) through a collaboration of six US national laboratories. The facility will produce pulsed beams of neutrons for use in scattering experiments by researchers from throughout the world. LBNL is responsible for the design and construction of the SNS Front End [1], which will produce a 52 mA, 2.5 MeV, 6% duty factor H- beam for injection into a 1 GeV linac. The Front End consists of several components: an ion source [2] and low energy beam transport line (LEBT) [3], an RFQ [4,5] and a medium energy beam transport line (MEBT) [6].

The RFQ resonant frequency is a function of both the cavity geometry and particularly the spacing of the vane tips. There is a frequency shift of approximately 1 MHz per 0.001 inch (25 microns) change in the tip-to-tip spacing. This dependence results in very tight machining tolerances on the individual vanes of ±0.0003 inch (8 microns) in order to achieve a final frequency which is within the range of the fixed slug tuners. During operation, a combination of RF power dissipated in the cavity walls and heat removal through the cooling passages will cause the cavity to distort and shift in frequency. By appropriately controlling the temperature of the RFQ cooling water continuously during operation, the cavity design frequency of 402.5 MHz will be maintained.

This paper will summarize the studies conducted to determine the RFQ frequency sensitivity to cooling water temperature changes. The predicted values will be compared to those obtained experimentally through testing of the completed prototype RFQ module (Figure 1). The operating scenario required to rapidly bring the system up to full RF power while maintaining the required frequency through cooling water control will also be presented.

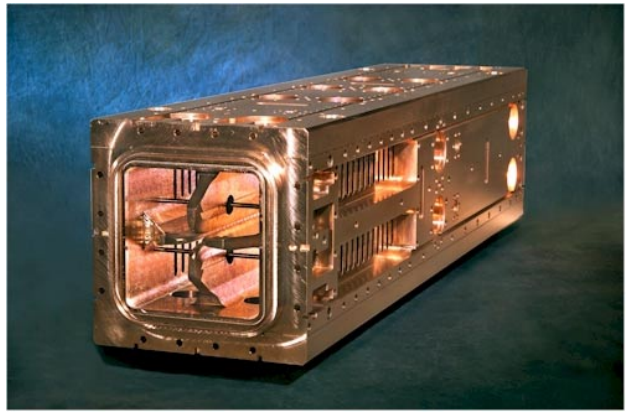

Figure 1: First completed SNS RFQ module.

## 2 RFQ DESCRIPTION

The SNS RFQ will accelerate an H- ion beam from 65 keV to 2.5 MeV over its 3.73 meter length. The 4 modules are constructed of C10100 oxygen free copper (OFE) with an outer layer of Glidcop AL-15. The OFE copper was selected due to its superior brazing characteristics and the Glidcop for its ability to maintain strength after brazing. The GlidCop is brazed to the outer surface of the OFE and covers the cooling passages which are milled into the back side of the copper vane piece. The vacuum seals for all penetrations (RF ports, tuners, vacuum ports and sensing loops) are recessed beyond the outer layer of GlidCop such that the braze is not exposed to the cavity vacuum. Also, since the cooling channels do not penetrate the ends of the modules, there are no water-to-vacuum joints in the entire system.

---

[*] This work is supported by the Director, Office of Science, Office of Basic Energy Sciences, of the U.S. Department of Energy under Contract No. DE-AC03-76SF00098.

For joining the vanes together, a zero-thickness brazing process has been selected in order to maintain the ±0.001 inch (25 microns) vane tip-to-vane tip tolerance. With this method, the joint surfaces are brought into close contact with Cusil wire braze alloy having been loaded into grooves in the joint surfaces prior to assembly. The alloy is spread throughout the joint during the braze cycle by means of capillary action. This technique permits the RFQ modules to be assembled and the cavity frequency measured prior to the braze cycle to allow for dimensional adjustments, if necessary.

The RF-induced thermal load on the cavity walls is removed by means of a dual temperature water cooling system. This setup allows fine tuning of the structure frequency in operation as well as during the RF power transient at start-up. A schematic of the RFQ cross section showing the OFE copper, Glidcop and cooling channel geometry is shown in Figure 2. The 12 outer wall channels are on a separately controlled water circuit from the 4 vane channels.

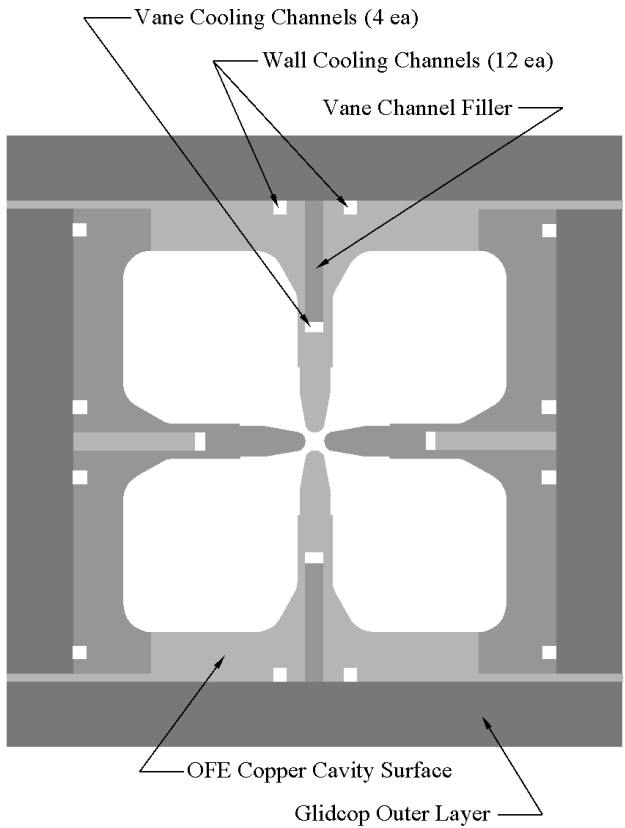

Figure 2: SNS RFQ cross section geometry.

## 3 FINITE ELEMENT MODELING

A finite element model of the RFQ has been developed using ANSYS and consists of a 3-D slice of one quadrant of the RFQ cross section. The surface nodes on either side of the slice are constrained to remain coplanar such that the longitudinal stresses are correctly calculated while allowing for overall thermal growth along the RFQ axis. This could not be achieved with 2-D plane strain elements which would over-constrain the model longitudinally and result in artificially high z-component compressive stresses. The loads and constraints applied to the model include RF heating on the cavity walls, external atmospheric pressure, convective heat transfer and water pressure on the cooling passage walls and boundary conditions imposed by symmetry constraints. With 18°C water in the vane channels and 24°C water in the cavity walls, the resulting temperature profile ranges between 25 and 29°C at full RF gradient with an average linear power density of 90 W/cm (Figure 3). The average power density on the outer wall is 1.7 W/cm$^2$. The model has also been used to calculate stresses and cavity wall displacements.

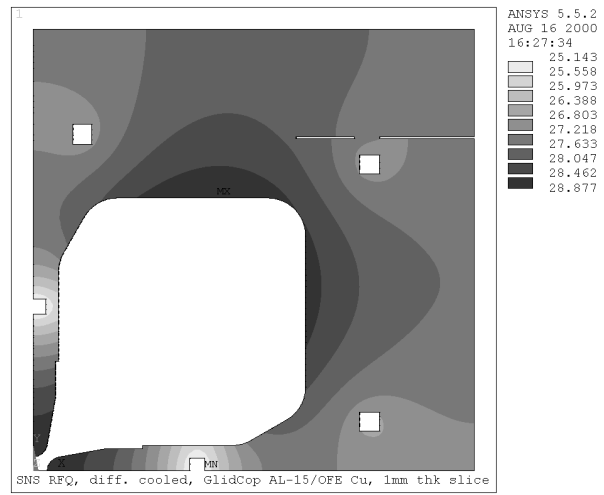

Figure 3: Predicted RFQ cavity wall temperature profile.

## 4 FREQUENCY SHIFT STUDIES

In order to predict the frequency shift of the RFQ cavity under various thermal conditions, a computer program was developed which combines the ANSYS displacement results with SUPERFISH calculations of frequency sensitivity. It was determined that the RFQ frequency shifts by –33 kHz for every 1°C rise in the vane cooling water. Preliminary measurements on the completed first RFQ module have yielded a value of –32 kHz/°C. This sensitivity to vane water temperature will be used to fine tune the RFQ frequency during operation based on sensing probe measurements. The calculated frequency shift for changes in the wall water temperature is +26 kHz/°C. For equal changes in the vane and wall water temperatures, the shift is -7 kHz/°C.

The calculations described above were based on nominal input temperatures for the vane and wall cooling channels. However, as the water flows from the inlet to outlet end of each RFQ module, its temperature will rise as it absorbs heat. Also, there will be a net heat transfer from the higher temperature wall water to the lower temperature vane water. The predicted values of 2.7°C rise in vane water temperature and 0.4°C rise in wall water temperature create a different cross section temperature profile at the outlet end of the RFQ module. The calculated frequency error due to the higher water temperatures is –80 kHz from the inlet to the outlet end of a 93 cm long module. This error is considered minor and can be corrected by adjusting the position of the fixed slug tuners along the length of the RFQ modules if necessary.

## 5 START-UP TRANSIENT

A series of transient analyses were performed using the same FEA model to determine the frequency performance of the system during ramp-up of the RF power. With 18°C water in the vanes and 24°C water in the walls and no heat on the cavity walls, the resonant frequency is 216 kHz higher than the nominal 402.5 MHz, outside the passband of the cavity. Setting the vane water at a higher temperature and the wall water at a lower temperature will bring the cavity frequency down. As the RF power is applied, the water temperatures are correspondingly adjusted towards their nominal operating values. However, the ramp-up rate for the RF power must be limited since the cooling systems cannot respond fast enough to keep the frequency error low. Figure 4 shows a comparison between the frequency shift caused by the water temperatures and that due to the cavity wall heat.

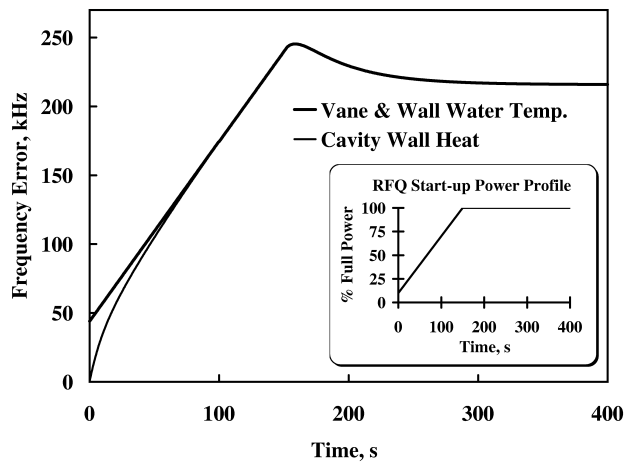

Figure 4: RFQ transient frequency error.

The error for wall heat is actually negative, but is plotted as positive for comparison purposes. The difference between these 2 curves is the net frequency error versus time. As shown in the inset, the power is ramped up from 10% to 100% of full power in 150 seconds. Initially, the water temperatures are adjusted at the highest rate possible until the 2 curves meet. This operating scheme results in an acceptable frequency error of 45 kHz at low power and drops off to less than 5 kHz within 60 seconds.

## 6 CONCLUDING REMARKS

The SNS RFQ resonant frequency will be regulated by dynamically adjusting the water temperature in the 4 cooling channels embedded in the vanes while holding the water temperature in the 12 wall channels constant. The theoretical frequency sensitivity to vane water temperature of –33 kHz/°C was confirmed by a measurement of –32 kHz/°C on the completed first RFQ module. Also, by adjusting both wall and vane water temperatures during start-up, the RF power can be increased to its full value in 150 seconds or less while holding the frequency error low enough to allow RF power transfer.

## 7 REFERENCES

[1] R. Keller for the Front-End Systems Team, "Status of the SNS Front-End Systems", EPAC '00, Vienna, June 2000.
[2] M.A. Leitner, D.W. Cheng, S.K. Mukherjee, J. Greer, P.K. Scott, M.D. Williams, K.N. Leung, R. Keller, and R.A. Gough, "High-Current, High-Duty-Factor Experiments with the RF Driven H- Ion Source for the Spallation Neutron Source", PAC '99, New York, April 1999, 1911-1913.
[3] D.W. Cheng, M.D. Hoff, K.D. Kennedy, M.A. Leitner, J.W. Staples, M.D. Williams, K.N. Leung, R. Keller, R.A. Gough, "Design of the Prototype Low Energy Beam Transport Line for the Spallation Neutron Source", PAC '99, New York, April 1999, 1958-1960.
[4] A. Ratti, R. DiGennaro, R.A. Gough, M. Hoff, R. Keller, K. Kennedy, R. MacGill, J. Staples, S. Virostek, and R. Yourd, "The Design of a High Current, High Duty Factor RFQ for the SNS", EPAC '00, Vienna, June 2000.
[5] A. Ratti, R.A. Gough, M. Hoff, R. Keller, K. Kennedy, R. MacGill, J. Staples, S. Virostek, and R. Yourd, "Fabrication and Testing of the First Module of the SNS RFQ", Linac '00, Monterey, August 2000.
[6] J. Staples, D. Oshatz, and T. Saleh, "Design of the SNS MEBT", Linac '00, Monterey, August 2000.